\documentstyle[12pt,epsf]{article}
\begin{document}


\pagestyle{empty}

\renewcommand{\thefootnote}{\fnsymbol{footnote}}
                                                  

\begin{flushright}
{\small
SLAC--PUB--7266\\
December 1996\\}
\end{flushright}
                
\begin{center}
{\bf\large
 Measurement of the $B^+$ and $B^0$ Lifetimes using Topological 
 Vertexing\footnote{Work supported in part by the
Department of Energy contract  DE--AC03--76SF00515.}}

\bigskip

The SLD Collaboration
\smallskip

Stanford Linear Accelerator Center, \\
Stanford University, Stanford, CA 94309\\
\medskip

\vspace{2.5cm}

{\bf\large
Abstract }
\end{center}

   The lifetimes of $B^+$ and $B^0$  
 mesons have been measured using
   a sample of 150,000 hadronic $Z^0$ decays
  collected by the SLD experiment at the SLC 
   between 1993 and 1995. 
   The analysis  
   reconstructs the decay length and charge of the $B$ meson using a novel 
   topological vertexing technique. 
   This method results in a high statistics sample of 6033 (3665) charged 
   (neutral) vertices with good charge purity. 
   A maximum likelihood fit procedure finds:   
   $\tau_{B^+}=1.67\pm0.07($stat$)\pm0.06$(syst) ps, 
   $\tau_{B^0}=1.66\pm0.08($stat$)\pm0.08$(syst) ps, 
   $\tau_{B^+}/\tau_{B^0} = 1.01^{+0.09}_{-0.08}($stat$)\pm0.05$(syst). 

\vspace{2cm}

\begin{center}

Submitted to {\sl Physical Review Letters}

\end{center}
\vfill

\renewcommand{\baselinestretch}{2}
\normalsize

\pagebreak
\pagestyle{plain}

%
%
%
  \def\iADEL{$^{(1)}$}
  \def\iBOL{$^{(2)}$}
  \def\iBU{$^{(3)}$}
  \def\iBRUN{$^{(4)}$}
  \def\iUCSB{$^{(5)}$}
  \def\iUCSC{$^{(6)}$}
  \def\iCIN{$^{(7)}$}
  \def\iCSU{$^{(8)}$}
  \def\iCOLO{$^{(9)}$}
  \def\iCOL{$^{(10)}$}
  \def\iFER{$^{(11)}$}
  \def\iFRA{$^{(12)}$}
  \def\iILL{$^{(13)}$}
  \def\iLBL{$^{(14)}$}
  \def\iMIT{$^{(15)}$}
  \def\iMASS{$^{(16)}$}
  \def\iMISS{$^{(17)}$}
  \def\iMOSC{$^{(18)}$}
  \def\iNAG{$^{(19)}$}
  \def\iOREG{$^{(20)}$}
  \def\iPAD{$^{(21)}$}
  \def\iPERU{$^{(22)}$}
  \def\iPISA{$^{(23)}$}
  \def\iRUT{$^{(24)}$}
  \def\iRAL{$^{(25)}$}
  \def\iSOGANG{$^{(26)}$}
  \def\iSOONG{$^{(27)}$}
  \def\iSLAC{$^{(28)}$}
  \def\iTENN{$^{(29)}$}
  \def\iTOH{$^{(30)}$}
  \def\iVAND{$^{(31)}$}
  \def\iWASH{$^{(32)}$}
  \def\iWISC{$^{(33)}$}
  \def\iYALE{$^{(34)}$}
  \def\dead{$^{\dag}$}
  \def\andgen{$^{(a)}$}
  \def\andper{$^{(b)}$}
%
%
\begin{center}
\mbox{K. Abe                 \unskip,\iNAG}
\mbox{K. Abe                 \unskip,\iTOH}
\mbox{T. Akagi               \unskip,\iSLAC}
\mbox{N.J. Allen             \unskip,\iBRUN}
\mbox{W.W. Ash               \unskip,\iSLAC$^\dagger$}
\mbox{D. Aston               \unskip,\iSLAC}
\mbox{K.G. Baird             \unskip,\iMASS}
\mbox{C. Baltay              \unskip,\iYALE}
\mbox{H.R. Band              \unskip,\iWISC}
\mbox{M.B. Barakat           \unskip,\iYALE
% ...
}
\it
  \vskip \baselineskip                   
  \centerline{(The SLD Collaboration)}   
  \vskip \baselineskip                   
%
%
%
  \iADEL
     Adelphi University,
     Garden City, New York 11530 \break
  \iBOL
     INFN Sezione di Bologna,
     I-40126 Bologna, Italy \break
  \iBU
     Boston University,
     Boston, Massachusetts 02215 \break
  \iBRUN
     Brunel University,
     Uxbridge, Middlesex UB8 3PH, United Kingdom \break
  \iUCSB
     University of California at Santa Barbara,
     Santa Barbara, California 93106 \break
  \iUCSC
     University of California at Santa Cruz,
     Santa Cruz, California 95064 \break
  \iCIN
     University of Cincinnati,
     Cincinnati, Ohio 45221 \break
  \iCSU
     Colorado State University,
     Fort Collins, Colorado 80523 \break
  \iCOLO
     University of Colorado,
     Boulder, Colorado 80309 \break
  \iCOL
     Columbia University,
     New York, New York 10027 \break
  \iFER
     INFN Sezione di Ferrara and Universit\`a di Ferrara,
     I-44100 Ferrara, Italy \break
  \iFRA
     INFN  Lab. Nazionali di Frascati,
     I-00044 Frascati, Italy \break
  \iILL
     University of Illinois,
     Urbana, Illinois 61801 \break
  \iLBL
     Lawrence Berkeley Laboratory, University of California,
     Berkeley, California 94720 \break
  \iMIT
     Massachusetts Institute of Technology,
     Cambridge, Massachusetts 02139 \break
  \iMASS
     University of Massachusetts,
     Amherst, Massachusetts 01003 \break
  \iMISS
     University of Mississippi,
     University, Mississippi  38677 \break
  \iMOSC
    Moscow State University,
    Institute of Nuclear Physics
    119899 Moscow, Russia    \break
  \iNAG
     Nagoya University,
     Chikusa-ku, Nagoya 464 Japan  \break
  \iOREG
     University of Oregon,
     Eugene, Oregon 97403 \break
  \iPAD
     INFN Sezione di Padova and Universit\`a di Padova,
     I-35100 Padova, Italy \break
  \iPERU
     INFN Sezione di Perugia and Universit\`a di Perugia,
     I-06100 Perugia, Italy \break
  \iPISA
     INFN Sezione di Pisa and Universit\`a di Pisa,
     I-56100 Pisa, Italy \break
  \iRUT
     Rutgers University,
     Piscataway, New Jersey 08855 \break
  \iRAL
     Rutherford Appleton Laboratory,
     Chilton, Didcot, Oxon OX11 0QX United Kingdom \break
  \iSOGANG
     Sogang University,
     Seoul, Korea \break
  \iSOONG
     Soongsil University,
     Seoul, Korea  156-743 \break
  \iSLAC
     Stanford Linear Accelerator Center, Stanford University,
     Stanford, California 94309 \break
  \iTENN
     University of Tennessee,
     Knoxville, Tennessee 37996 \break
  \iTOH
     Tohoku University,
     Sendai 980 Japan \break
  \iVAND
     Vanderbilt University,
     Nashville, Tennessee 37235 \break
  \iWASH
     University of Washington,
     Seattle, Washington 98195 \break
  \iWISC
     University of Wisconsin,
     Madison, Wisconsin 53706 \break
  \iYALE
     Yale University,
     New Haven, Connecticut 06511 \break
  \dead
     Deceased \break
  \andgen
     Also at the Universit\`a di Genova \break
  \andper
     Also at the Universit\`a di Perugia \break
\rm
%

\end{center}


\pagebreak

 The spectator model predicts that the lifetime of a 
 heavy hadron depends
 upon the properties of the constituent weakly decaying heavy quark $Q$ and 
 is independent of the remaining, or spectator, quarks in the hadron.
 This model fails for the charm hadron system for which the
 lifetime hierarchy 
 $\tau_{D^+} \sim 2\tau_{D_s^+} \sim 2.5\tau_{D^0} 
 \sim 5\tau_{\Lambda_c^+}$ is observed.
 Since corrections to the spectator model are predicted to scale with
  $1/m_Q^2$ the $B$ meson lifetimes are expected to
  differ by less than 10\% \cite{bigi}. 
 Hence a measurement of the $B^+$ and $B^0$ lifetimes 
 provides a test of this prediction.
In addition, the specific $B$ meson lifetimes are needed to determine the  
 element $V_{cb}$ of the CKM matrix. 

The analysis is performed on the 1993-5 data sample of 150,000
$Z^0$ decays collected by SLD 
at the SLC.
 The excellent 3D vertexing capabilities of SLD 
 are exploited 
 with a novel topological vertexing technique \cite{zvnim} to
 identify $B$ hadron vertices
 produced in hadronic $Z^0$ decays with high efficiency.
 The decay length is measured using the
reconstructed vertex location while the $B$ hadron charge is determined from
the total charge of the tracks associated with the vertex.
 This inclusive technique has the advantage of very efficient
  $B$ vertex reconstruction since essentially all $B$ decays are used.

The components of the SLD 
 utilized by this analysis are the Central
Drift Chamber (CDC)\cite{rbrb}
for charged track reconstruction and momentum measurement and the CCD pixel
Vertex Detector (VXD)\cite{rbrb}
 for precise position measurements near the interaction
point. These systems are immersed in the 0.6 T field of the SLD solenoid.
Charged tracks reconstructed in the CDC are linked with pixel clusters in the
VXD by extrapolating each track and selecting the best set of associated
clusters\cite{rbrb}.
For a typical track from the primary vertex or heavy hadron decay, the total
efficiency of reconstruction in the CDC and linking to a correct set of VXD
hits is 94\% for the region $\left|\cos\theta\right|<0.74$.
The track impact parameter resolutions at high momentum
are 11~$\mu$m and 38~$\mu$m in the $r\phi$ and $rz$ projections
respectively ($z$ points along the beam direction),
while multiple scattering contributions are
$70 \,\mu$m~$/(p\,{\rm sin}^{3/2}\theta)$ in both projections (where the
momentum $p$ is expressed in GeV/c).

The centroid of the micron-sized SLC Interaction Point (IP) in the $r\phi$
plane is reconstructed with a
measured precision of $\sigma_{IP} = (7\pm2)\, \mu$m using tracks in sets of
$\sim30$ sequential hadronic $Z^0$ decays. The median $z$ position of tracks
at their point of closest approach to the IP in the $r\phi$ plane is used to
determine the $z$ position of the $Z^0$ primary vertex on an event-by-event
basis.  A precision of $\sim52\,\mu$m~\cite{rbrb} on this
quantity is estimated using $Z^0 \rightarrow b\overline{b}$
  Monte Carlo simulation.

 The simulated events are generated using JETSET 7.4 \cite{jetset}.
 The $B$ meson decays are simulated using the CLEO $B$ decay model
 \cite{CLEO-QQ} tuned to reproduce the spectra and
 multiplicities
 of charmed hadrons, pions, kaons, protons and leptons as measured at the
 $\Upsilon$(4S) by ARGUS and CLEO \cite{argcl,muheim}.
 The branching fractions
 of the charm hadrons are tuned to the existing measurements
 \cite{PDG94}. The $B$ mesons and baryons are generated
 with a lifetime of 1.55 ps and 1.10 ps respectively, while the $b$-quark
 fragmentation follows the Peterson {\em et al.} parametrization
 \cite{Peterson}.
 The SLD detector is simulated using GEANT 3.21 \cite{geant}.

  Hadronic $Z^0$ event selection requires at least 7
 CDC tracks  which pass within 5~cm of the IP in $z$ at the point
 of closest approach to the beam and which have
 momentum transverse to the beam direction $p_T>$200~MeV/$c$.
 The sum of the energy of the charged tracks passing these cuts
 must be greater than 18~GeV.
 These requirements remove background from $Z^0 \to
 l^+ l^-$ events and two-photon interactions. In addition, the
 thrust axis determined from energy clusters in the calorimeter must
 have $\left|\cos\theta\right|<0.71$, within the acceptance of the
 vertex detector.
 These requirements yield a sample
 of $\sim 96,000$ hadronic $Z^0$ decays.

  Good quality tracks used for vertex finding must have a CDC hit
   at a radius$<$39~cm, and have
  $\geq$40 hits to insure that
the lever arm provided by the CDC is appreciable.
The CDC tracks must have $p_T>$400~MeV/$c\:$ and 
extrapolate to within 1~cm of the IP in $r\phi$
and within 1.5~cm in
$z$ to eliminate tracks which arise 
from interaction with the detector material.
The fit of the track must satisfy $\chi^2/$d.o.f.$<5$.
At least one good VXD link is required, and the combined
CDC/VXD fit must also satisfy $\chi^2/$d.o.f.$<5$.

     The topological vertex reconstruction is applied separately to
 the tracks in each hemisphere (defined with respect to the 
 event thrust axis).
 This analysis is the first application
 of the algorithm which is described in detail
 in Ref. \cite{zvnim} and summarized here.
    The vertices are reconstructed in 3D co-ordinate space
  by defining a vertex function $V({\bf r})$ at each
  position ${\bf r}$. The helix
 parameters for each track $i$ are used to describe
  the 3D track trajectory as a Gaussian tube $f_i({\bf r})$, where the
  width of the tube is the uncertainty in the measured track location
  close to the IP.
  A function $f_0({\bf r})$ is used to describe the location and
  uncertainty of the IP.
 $V({\bf r})$ is defined as a function of $f_0({\bf r})$ and
  the $f_i({\bf r})$ such that it is
   small in regions where fewer than two tracks (required
  for a vertex) have significant $f_i({\bf r})$, and large in
 regions of high track multiplicity.
  Maxima are found in $V({\bf r})$ and
  clustered into resolved spatial regions.
  Tracks are associated with these regions to form a set of
  topological vertices.

 The efficiency for reconstructing $B$ hadron decay vertices
 is 80\% for true decay lengths greater than $3$mm,
 as estimated by the simulation. The efficiency
 falls at shorter decay length as it becomes harder to resolve
 the secondary vertex from the IP.
 The efficiency for reconstructing at least one secondary vertex
  is $\sim 50$\% in $b$ hemispheres, $\sim 15$\% in
 charm hemispheres and $\sim 3$\% in light quark hemispheres. The
 efficiency for reconstructing
 more than one secondary vertex is $\sim 5$\%
  in $b$ hemispheres.
 For  hemispheres containing secondary vertices,
 the `seed' vertex is chosen to be the one
  furthest from the IP.
  Vertices consistent with a $K^0_s \to \pi^+\pi^-$ decay
  are excluded from the seed
 vertex selection and the two tracks are discarded.

 A vertex axis is formed by a straight line joining the IP
  to the seed vertex.
 The 3D distance of closest approach of a track to the vertex axis, T,
  and the distance from the IP along the vertex
 axis to this point, L, are calculated for all quality tracks.
  Monte Carlo studies show that tracks which are not directly associated
  with the seed vertex
  but which pass T$<0.1$~cm and L$/$D$>0.3$ (where D is the distance
  from the IP to the seed vertex) are more
 likely to have been produced by the $B$ decay sequence than
  to have an alternative origin.
 Hence such tracks are added to the set of
  tracks in the seed vertex to form the candidate $B$ decay vertex,
  containing tracks from both the $B$ and cascade
  $D$ decays.  The distance
  from the IP to the location determined by fitting this set of tracks
 to a common vertex is the reconstructed decay length.
 Since the purity of the $B$ charge
 reconstruction is lower for decays close to the IP, where tracks are
 more likely to be wrongly assigned, decay lengths are required
 to be $>1$~mm.

\begin{figure}[htb]
\centering
\epsfxsize11cm
\leavevmode
\epsfbox{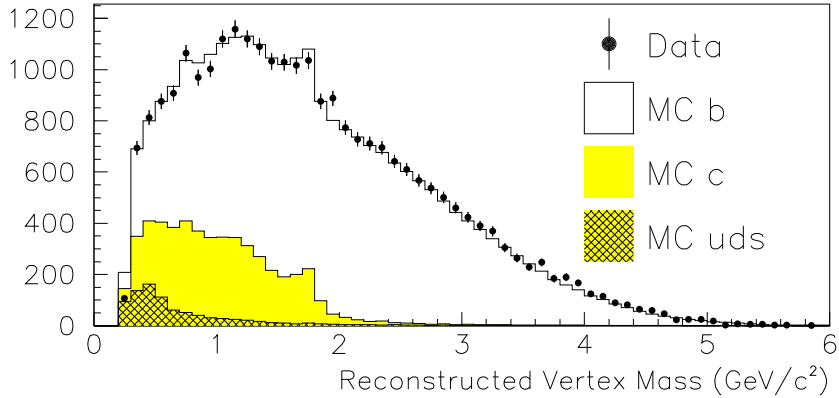}
\caption{Mass of reconstructed vertex for 
 data (points) and Monte Carlo (histogram).}  
\label{hadmassp}
\end{figure}

 The lifetime measurement relies on the ability to separate
  $B^+$ and $B^0$ decays by making use of the  vertex charge.
   Monte Carlo studies show that
 the purity of the charge reconstruction
 is more likely to be eroded by losing
 tracks from the $B$ decay chain through track selection
 inefficiencies and track mis-assignment than by gaining mis-assigned
 tracks originating from the primary or other background to the $B$
 decay.
 Furthermore, the decays which are missing some $B$ tracks
  tend to have
 lower vertex mass as well as lower charge purity.
 (The mass is calculated by assuming each
 track associated with the vertex has the mass of a pion.)
 Hence the vertex mass is required
 to be $>2$ GeV/$c^2$ to improve the charge reconstruction purity.
 In addition, the mass distribution (see Fig.~\ref{hadmassp})
 shows that a large
 fraction of the charm and light flavor
contamination is eliminated by
 this cut.  A sample of 9719 candidate $B$ decay vertices
 remains, with a mean 
track multiplicity of 5.0.

\begin{figure}[htb]
\centering
\epsfxsize9cm
\leavevmode
\epsfbox{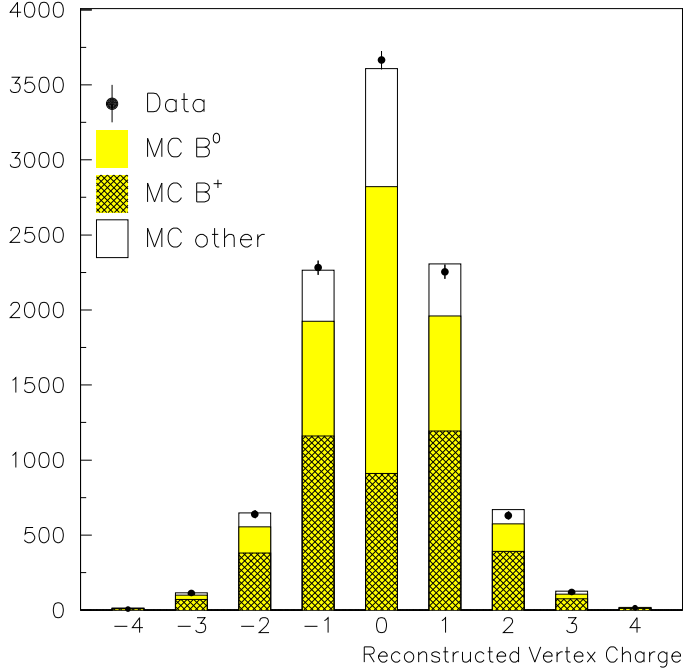}
\caption{\label{hadcharge}Reconstructed vertex charge for 
 data (points) and Monte Carlo (histogram).}  
\end{figure}

  To improve the $B$ hadron charge reconstruction,
 tracks which
  fail the initial selection but have $p_T>$ 200 MeV/$c$ and
  $\sqrt{\sigma_{r\phi}^2 + \sigma_{rz}^2}<700\,\mu$m,
 where $\sigma_{r\phi}$ ($\sigma_{rz}$)
 is the uncertainty in the track position
 in the $r\phi$ ($rz$) plane
 close to the IP, are
  considered as decay track candidates.
    The charge of these tracks which
 pass the cuts
 T$<\,0.1$~cm and L$/$D$>0.3$ is added to the $B$ decay charge. On 
 average, 0.5 of these lower quality tracks pass these criteria in
 $b$ hemispheres.

 Fig.~\ref{hadcharge} shows a comparison of the reconstructed
charge between data and Monte Carlo.
   The charged sample consists of 6033 vertices with vertex
  charge equal to $\pm $ 1,2 or 3,
  while the neutral sample consists
  of 3665 vertices with charge equal to 0.
Monte Carlo studies
indicate that the charged sample is 97.8\% pure in $B$ hadrons
consisting of
52.8\% $B^+$, 32.1\% $B^0$, 8.6\% $B_s^0$, and 4.3\% $B$ baryons.
(Charge conjugation is implied throughout this paper.) 
 Similarly, the neutral sample is 98.3\% pure in $B$ hadrons
consisting of
25.3\% $B^+$, 52.9\% $B^0$, 13.9\% $B_s^0$ and 6.2\% $B$ baryons.
 The statistical precision of the measurement depends on the
 separation between the $B^+$ and $B^0$ in these samples.

\begin{figure}[htb]
\centering
\epsfxsize12cm
\leavevmode
\epsfbox{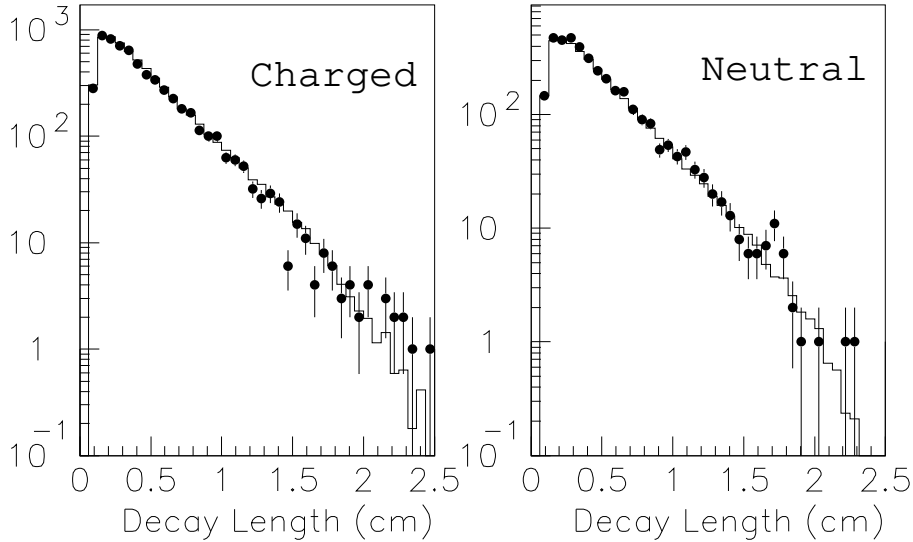}
\caption{Decay length distributions for 
  data (points) and best fit Monte Carlo (histogram).}
\label{topdklen}
\end{figure}

 The lifetimes are extracted from the decay length distributions, 
  shown in Fig.~\ref{topdklen}, for the
selected charged and neutral $B$ decay samples 
 using a binned maximum likelihood technique,
in which the Monte Carlo $B^+$ and $B^0$ decays are weighted to yield 
decay length distributions for
 varying $B^+$ and $B^0$ lifetimes that are
 compared to the data \cite{semil}. 
A two parameter fit (over the range 1~mm to 25~mm) yields
 lifetimes of
   $\tau_{B^+}=1.67\pm0.07$ ps and
   $\tau_{B^0}=1.66\pm0.08$ ps,
   with a ratio of $\tau_{B^+}/\tau_{B^0}$
   $ = 1.01^{+0.09}_{-0.08}$ and
 a combined $\chi^2/$d.o.f. = 90.0/76.

Table~\ref{systerrs} summarizes the systematic errors on the $B^+$ and
$B^0$ lifetimes and their ratio.
To account for a 
 discrepancy between data and Monte Carlo
in the fraction of tracks passing the selection criteria,
a 4\% tracking efficiency correction 
with dependence on track momenta and angles
 is applied to the simulation \cite{rbrb}.
 The corrected Monte Carlo is used in the lifetime fits, with the 
 effect of the entire correction 
taken as the systematic error.  The decay length 
 distribution of the smaller neutral sample, and hence the 
 measured $B^0$ lifetime,
 is perturbed more than the charged sample by this conservative
 estimate of the tracking efficiency uncertainty. 
 The uncertainty due to tracking resolution, mainly due to remaining
vertex detector misalignments in the {\it rz} plane,
 is estimated by applying  to the
Monte Carlo track {\it rz} impact parameter $\phi$
 dependent systematic shifts   of up to
20 ${\mu}$m and a random Gaussian smear with
 $\sigma=20{\mu}\mbox{m}/\sin\theta$.
 The total effect of
 this correction is again assigned as a systematic error.
We have also made cross checks by performing the lifetime fits for $B$ decay
candidates in different $\phi$ regions and different data taking time periods
separately. The results are found to be consistent within statistics.

\begin{table}[htb]
\begin{center}
\caption{\label{systerrs} Summary of systematic uncertainties in 
         the $B^+$ and $B^0$ lifetimes and their ratio.}

\begin{tabular}{lcccc}
\\
\hline
Systematic Error  & & $\Delta\tau_{B^+}$ 
                    & $\Delta\tau_{B^0}$ 
                    & $\Delta\frac{\tau_{B^+}}{\tau_{B^0}}$  \\
  & & (ps) & (ps) & \\ 
   \hline 
 & \multicolumn{3}{c}{Detector Modeling} \\ \hline
 Tracking efficiency& & 0.011  & 0.035 & 0.028 \\
 Tracking resolution& & 0.012  & 0.011 & 0.010 \\
\hline
 & \multicolumn{3}{c}{Physics Modeling} \\ \hline
$b$ fragmentation  & 0.700 $\pm$ 0.011 & 0.035 & 0.037 & 0.005 \\
BR($B \to DX$)   &  & 0.010    & 0.012  & 0.010    \\
BR($B \to D\overline{D}X$)   & 0.15 $\pm$ 0.05 & 0.006 & 0.006 & 0.006 \\
$B$ decay multiplicity & 5.3 $\pm$ 0.3 & 0.016    & 0.012  & 0.003    \\
$B_s^0$ fraction  & 0.115 $\pm$ 0.040 & 0.012    & 0.004  & 0.005  \\
$B$ baryon fraction & 0.072 $\pm$ 0.040 & 0.013    & 0.039  & 0.017  \\
$B_s^0$ lifetime & 1.55 $\pm$ 0.10 ps  & $<$.003    & 0.025  & 0.016  \\
$B$ baryon lifetime & 1.10 $\pm$ 0.08 ps & $<$.003  & 0.006  & 0.004  \\
$D$ decay multiplicity&  & 0.011 & 0.006 & 0.010 \\
$D$ decay $K^0$ yield& & 0.005   & 0.020   & 0.010   \\
\hline
 & \multicolumn{3}{c}{Monte Carlo and Fitting} \\ \hline
Fitting systematics & & 0.024    & 0.013 & 0.022   \\
MC statistics      & & 0.012    & 0.013 & 0.015   \\
\hline
TOTAL            &   & 0.055    & 0.078 & 0.050     \\
\hline
\end{tabular}
\end{center}
\end{table}

The physics modeling systematic uncertainties were determined as follows.
The mean fragmentation energy $<\!\! x_E\!\!>$ of the $B$ hadron 
 \cite{bfrag} 
 and the shape of the $x_E$ distribution \cite{Bowler} were varied.
 Since the fragmentation is assumed to be identical for the $B^+$ and
 $B^0$ mesons, this uncertainty has little effect on the lifetime ratio.
The four branching fractions
for $B^+/B^0\rightarrow \overline{D^0}/D^- X$
were varied by twice the
uncertainty in the current world average for
 $B\rightarrow \overline{D^0}/D^- X$ \cite{muheim}. 
The fraction of $B^+/B^0$ decays producing a $D\overline{D}$ pair was also
 varied.
The average $B^+$ and $B^0$ decay multiplicity was varied by
 $\pm 0.3$ tracks \cite{bmult} in an anticorrelated manner.
Uncertainties in the $B_s^0$ and $B$ baryon lifetimes and production
 fractions mostly affect the $B^0$ lifetime since 
the  neutral $B_s^0$ and $B$ baryon are a
 more significant background
  for the $B^0$ decays.
The systematic errors due to uncertainties in charmed meson decay
topology were estimated by changing the Monte Carlo
 $D$ decay charged multiplicity and $K^0$ production
according to the uncertainties in experimental measurements \cite{MK3DDECAY}.
The effect of varying the lifetime of charm hadrons
  ($D^+$, $D^0$, $D_s$, $\Lambda_c$), as well as their momentum spectra in
 the $B$ decay rest frame was found to be negligible.

 The fitting uncertainties were determined
by varying the bin size used in the decay length distributions,
and by modifying the cuts on the minimum (0--2~mm) and maximum
 (12--25~mm) decay lengths
used in the fit. Fit results are consistent within statistics
for these variations, but a systematic error is conservatively assigned
using the RMS variation of the results.

In summary, from 150,000 $Z^0$ decays
 collected by SLD between 1993 and 1995,
 the $B^+$ and $B^0$ lifetimes have been measured using a novel 
 topological technique.
The analysis isolates 9698 $B$ hadron candidates with good
 charge purity
and determines the lifetimes of $B^+$ and $B^0$ mesons to be
$\tau_{B^+} = 1.67\pm0.07(\mbox{stat})\pm0.06(\mbox{syst})\mbox{~ps}$,
$\tau_{B^0} = 1.66\pm0.08(\mbox{stat})\pm0.08(\mbox{syst})\mbox{~ps}$,
with a ratio
$\frac{\tau_{B^+}}{\tau_{B^0}}=
   1.01^{+0.09}_{-0.08}(\mbox{stat})\pm0.05(\mbox{syst})$.
 Combining this measurement with that obtained using a complementary
 analysis based on 
  semileptonic decays \cite{semil}, taking into account correlated 
  statistical and systematic errors, yields the following SLD averages:

\begin{eqnarray}
\tau_{B^+} & = &
  1.66\pm0.06(\mbox{stat})\pm0.05(\mbox{syst})\mbox{~ps}, \nonumber \\
\tau_{B^0} & = &
  1.64\pm0.08(\mbox{stat})\pm0.08(\mbox{syst})\mbox{~ps}, \nonumber \\
\frac{\tau_{B^+}}{\tau_{B^0}} & = & 
   1.01\pm0.07(\mbox{stat})\pm0.06(\mbox{syst}). \nonumber
\end{eqnarray}

 These results are consistent with the
expectation that the $B^+$ and $B^0$ lifetimes are nearly equal
 and have a statistical accuracy among the best of the current
  measurements \cite{world}.

\noindent

        We thank the personnel of the SLAC accelerator department and
the technical staffs of our collaborating institutions for their outstanding
efforts.

\enddocument